\documentclass[11pt]{article}

\begin{document}
\begin{center}
{\Large\bf Path Integration in Conical Space}\\[5mm]

Akira Inomata\\

{\it Department of Physics, State University of New York at Albany \\
1400 Washington Avenue, Albany, NY 12222, USA}\\

and\\

Georg Junker\footnote{\noindent Corresponding author: Tel/fax: +49 89 32006231;
e-mail: gjunker@eso.org}
\\

{\it European Organization for Astronomical Research in the Southern
Hemisphere\\
Karl-Schwarzschild-Strasse 2, D-85748 Garching, Germany}\\
\end{center}

{\small
{\bf Abstract} Quantum mechanics in conical space is studied by the
path integral method. It is shown that the curvature effect gives rise
to an effective potential in the radial path integral. It is further
shown that the radial path integral in conical space can be reduced
to a form identical with that in flat space when the discrete angular
momentum of each partial wave is replaced by a specific non-integral
angular momentum. The effective potential is found proportional
to the squared mean curvature of the conical surface embedded in
Euclidean space. The path integral calculation is compatible with the
Schr\"odinger equation modified with the Gaussian and the mean
curvature.
\\[5mm]
{\bf PACS numbers and keywords}:\\
61.72Lk - Linear defects: dislocations, disclinations\\
98.80.Cq - Cosmic strings\\
03.65.-w Quantum mechanics: Path integral
}

\section{Introduction}
In recent years there has been considerable interest in quantum
mechanics in the field of topological defects (see, e.g., \cite{KF}).
Although the notion of defects in physics was originally associated with
crystalline irregularities, it has been extended to more general
topological structures such as entangled polymers, liquid crystals,
magnetic vortices, anyons, monopoles, cosmic strings, domain walls, and
so on. Standard approaches to particle-defect interactions in quantum
mechanics are to solve appropriate Schr\"odinger equations with relevant
boundary conditions. The space surrounding a defect is often
characterized by torsion and curvature. The torsion may be globally
treated through boundary conditions in connection with the topological
nature of the defect as in the case of the Aharonov-Bohm effect, whereas
the information of curvature would have to be fully contained in the
Schr\"odinger equation. In curved space, the Schr\"odinger equation is
usually expressed in the form,
\begin{equation}
\left\{-\frac{\hbar^2}{2M}\Delta + V_{c}({\bf r}) + V({\bf r})\right\}\psi ({\bf r}) =
i\hbar\frac{\partial}{\partial t}\psi ({\bf r})
\end{equation}
where $\Delta $ is the Laplace-Beltrami operator, $V_{c}({\bf r})$ is
the so-called curvature term and $V({\bf r})$ is any external potential.
Historically, Podolsky \cite{Pod} defined the Schr\"odinger equation in
curved space without the curvature term, namely $V_{c}({\bf r})=0$.
Comparing with the path integral formulation, DeWitt \cite{BdW} asserted
that the curvature term is proportional to the Ricci scalar curvature,
that is, $V_{c}({\bf r})=g\hbar^{2}R({\bf r})$ where $g$ is a constant.
Moreover, Jensen and Koppe \cite{JK}, and da Costa \cite{daC} argued, if
the particle in question is constrained to move on a two-dimensional
curved surface, the curvature term must be give by
\begin{equation}
V_{c}({\bf r})=\frac{\hbar^{2}}{2M}\left\{K({\bf r}) - H^{2}({\bf
r})\right\} \label{VC}
\end{equation}
where $K({\bf r})$ is the Gaussian curvature and $H({\bf r})$ is the
mean curvature of the curved surface.\footnote{
For more recent and more general discussions, see \cite{DMO}-\cite{FC}.}

In a recent paper \cite{IJR}, we have studied quantum mechanics in the
field of a dispiration (a combined structure of a dislocation and a
disclination), and have pointed out that the path integral calculation
leads to a result different from that of the Schr\"odinger equation in
the Podolsky form with no curvature term. The purpose of the present
paper is to show that for a particle moving in conical space the path
integral calculation is consistent with the Schr\"odinger equation only
when the curvature term is of the Jensen-Koppe form (\ref{VC}).
We find that the curvature effect of the conical surface gives rise to an
effective potential in the radial path integral and that the effective
potential is proportional to the squared mean curvature of the conical
surface.

\section{Conical space}

We consider a two-dimensional curved space with metric,
\begin{equation}
dl^{2}=dr^{2} + \sigma ^{2}r^{2}d\theta ^{2},    \label{dl}
\end{equation}
where $0 \leq r $ and $\theta \in [0, 2\pi )$, $\sigma $ being a
real parameter. If $|\sigma | \leq 1$, this space may be realized as a
conical surface when imbedded in the three-dimensional Euclidean
space with
\begin{equation}
\begin{array}{ll}
x= \sigma r\cos \theta & ~ \\
y= \sigma r\sin \theta & ~ \\
z= \sqrt{1 - \sigma ^{2}}\,r. & ~ \end{array}
\end{equation}
Obviously the limiting case where $\sigma =1$ corresponds to the
two-dimensional flat space with $z=0$.

The medium around an axial wedge dislocation in solid may
be characterized by such a space, for which $\sigma $ is related to
the deficit angle $\gamma $ via $\sigma = 1 - \gamma /2\pi $. See, e.g.,
\cite{IJR,PS}. Another example is the space surrounding a massive cosmic
string with a linear energy density $\eta $, given in a weak field
approximation, for which $\sigma = 1 -4G \eta $ where $G$ is the
Newtonian gravitational constant \cite{Vil}. The conical topology also
occurs in (2+1)-dimensional Einstein gravity with localized masses
\cite{DJt}. In this connection, the Schr\"odinger equation in conical
space has been further discussed in \cite{DJ} - \cite{FM}.

In the present paper, however, we study quantum mechanics in the space
with metric (\ref{dl}) by path integration. Specifically we carry
out path integration for a particle with mass $M$ moving in the conical
space with metric (\ref{dl}) under the influence of a two-dimensional
central potential $V(r)$. The Lagrangian for the particle is
\begin{equation}
L = \frac{M}{2}\left(\dot{r }^{2} + \sigma ^{2}r^{2}\dot{\theta
}^{2}\right) - V(r).
\end{equation}
Although the potential $V(r)$ will appropriately be chosen later for
explicit calculation, we assume that it contains a short-ranged
repulsive part. In fact, the Hamiltonian describing a free motion in a
conical space has a one-parameter family of self-adjoint extensions and
requires a careful definition of boundary conditions imposed at
$r=0$. Such boundary conditions are to be determined by the physics near
$r=0$ \cite{KS}. We circumvent such a singularity problem at $r=0$ by
assuming an appropriate short-ranged repulsive potential.

\section{The propagator}

What we wish to calculate by path integration is the propagator or
Feynman's kernel for the aforementioned system.
Feynman's path integral for the propagator may be given in the
time-sliced form \cite{Feyn},
\begin{equation}
K({\bf r}'', {\bf r}'; \tau )=\lim_{N \rightarrow \infty }\int
\prod_{j=1}^{N-1} d^2{\bf r}_{j}\,\prod_{j=1}^{N}K({\bf r}_{j}, {\bf
r}_{j-1}; \varepsilon ) \label{K}
\end{equation}
where the time interval $\tau =t''-t'$ is sliced into $N$ short-time
intervals $\varepsilon =\tau /N$. The short time propagator is given by
\begin{equation}
K({\bf r}_{j}, {\bf r}_{j-1}; \varepsilon )
= \frac{M\sigma }{2\pi i\hbar \varepsilon } \,e^{(i/\hbar)S_{j}}
\label{SK}
\end{equation}
with a short-time action,
\begin{equation}
S_{j}=\int_{t_{j-1}}^{t_{j}}L\,dt = \frac{M}{2\varepsilon }
\Delta {\bf r}_{j}^{2} - V(r_{j})\varepsilon.
\end{equation}
The amplitude of the short-time propagator (\ref{SK}) has been
determined by the condition,
\begin{equation}
\lim_{\varepsilon \rightarrow 0}K({\bf r}_{j}, {\bf r}_{j-1}; \varepsilon )
=\delta ({\bf r}_{j} - {\bf r}_{j-1}).
\end{equation}

Here we intend to carry out path integration for (\ref{K}) in polar
coordinates \cite{EG,PI,BJ} using the relations
\cite{IJR},
\begin{equation}
\prod_{j=1}^{N-1} d^2{\bf r}_{j} =\prod_{j=1}^{N-1} r_{j}dr_{j}\,d\theta _{j},
~~~~~~\Delta {\bf r}_{j}^{2}=\Delta r_{j}^{2} + 2\sigma
^{2}\hat{r}_{j}^{2}(1 - \cos \Delta \theta _{j})
\end{equation}
where $\hat{r}_{j}=(r_{j}r_{j-1})^{1/2}$. The angular integration is easily
performed by utilizing the generating function for the modified Bessel
function,
\begin{equation}
\exp\left\{z \cos \Delta \theta _{j}\right\}=\sum_{m=-\infty }^{\infty }
e^{im\Delta \theta _{j}}I_{m}(z),
\end{equation}
valid for any complex number $z$. As a result of the successive angular
integrations, the propagator (\ref{K}) can be expressed in the form of
the partial wave expansion,
\begin{equation}
K({\bf r}'', {\bf r}' ; \tau )=\frac{1}{2\pi }\sum_{m=-\infty }^{\infty
}e^{im(\theta ''-\theta ')}R_{m}(r'', r'; \tau ).
\label{KR}
\end{equation}
The radial propagator for the $m$-th partial waves still remains to be
path-integrated by
\begin{equation}
R_{m}(r'', r'; \tau )=\lim_{N\rightarrow \infty }\int \prod_{j=1}^{N-
1}\,r_{j}\,dr_{j}\,\prod_{j=1}^{N}R_{m}(r_{j}, r_{j-1}; \varepsilon )
\label{RK}
\end{equation}
with the short-time radial propagator
\begin{equation}
R_{m}(r_{j}, r_{j-1}; \varepsilon )= \frac{M}{i\hbar\varepsilon }
\exp\left\{\frac{i}{\hbar}\left[\frac{M}{2\varepsilon }(r_{j}^{2} +
r_{j-1}^{2}) - V(r_{j})\varepsilon \right]\right\}{\bf I}_{m}^{\sigma
}(\hat{r}_{j}) \label{SRK}
\end{equation}
where
\begin{equation}
{\bf I}_{m}^{\sigma }(\hat{r}_{j})=\sigma
\exp\left\{\frac{iM}{\hbar\varepsilon }(\sigma ^{2}-
1)\hat{r}_{j}^{2}\right\}I_{m}\left(\frac{M\sigma ^{2}}{i\hbar
\varepsilon }\hat{r}_{j}^{2}\right). \label{bfI}
\end{equation}
Notice that the parameter $\sigma $ appears in (\ref{SRK}) only though the
function (\ref{bfI}) and that
\begin{equation}
{\bf I}_{m}^{1}(\hat{r}_{j})
=\lim_{\sigma \rightarrow 1}{\bf I}_{m}^{\sigma }(\hat{r}_{j})
=I_{m}\left(\frac{M}{i\hbar \varepsilon }\hat{r}_{j}^{2}\right).
\label{bfI1}
\end{equation}
In the limit $\sigma \rightarrow 1$, the radial propagator (\ref{RK})
becomes that in flat space. Accordingly, (\ref{bfI1}) is the modified
Bessel function to appear in a flat space path integral. Hence the
effect of the deviation from the flat space path integral is all
contained in (\ref{bfI}). In what follows, using the asymptotic
recombination technique, we shall show that the deviation from the flat
space gives rise to an effective potential in the radial path integral.

\section{The curvature effect}

A very useful calculation method for a polar coordinate path integral
is the asymptotic recombination technique \cite{IKG}, to
which basic is the one-term asymptotic formula of the modified Bessel
function, originally employed by Edwards and Gulyaev \cite{EG},
\begin{equation}
I_{\nu }(z) \sim \frac{1}{\sqrt{2\pi z}}\,\exp\left\{z -
\frac{1}{2z}\left(\nu ^{2} - \frac{1}{4}\right)\right\}, \label{asy}
\end{equation}
valid for large $|z|$ with Re$z > 0$. In a polar coordinate path
integral, the complex variable $z$ is of the form $z=Mar^{2}/(i\hbar
\varepsilon )$ where $a$ is a positive real constant. In order to
justify the use of the asymptotic formula, it is necessary to assume
that $M/\hbar$ has a small positive imaginary part Im$ (M/\hbar) > 0$. The same
trick has been used in the standard analytic continuation procedure to
obtain a well-defined Feynman path integral. Of course, in this case,
$|z|$ is large when $\varepsilon $ is small. It is clear that the
asymptotic formula (\ref{asy}) is valid insofar as it is used for a
short-time Feynman path integral.

Substituting (\ref{asy}) with $z=M\sigma ^{2}\hat{r}^2_{j}/(i\hbar
\varepsilon )$ into (\ref{bfI}), combining the exponential factor of
(\ref{bfI}) with (\ref{asy}), and rearranging terms, we find the result,
\begin{equation}
{\bf I}_{m}^{\sigma }(\hat{r}_{j}) \sim \exp\left\{-
\frac{i}{\hbar}V_{eff}(\hat{r}_{j})\varepsilon \right\}\,I_{m/\sigma
}\left(\frac{M}{i\hbar\varepsilon }\hat{r}_{j}^{2}\right) \label{bfI2}
\end{equation}
where
\begin{equation}
V_{eff}(\hat{r}_{j})=- \frac{\hbar^{2}}{2M}\frac{1 - \sigma
^{2}}{4\sigma ^{2}\hat{r}_{j}^{2}}.  \label{effV}
\end{equation}
Correspondingly, the short-time radial propagator (\ref{SRK}) can be written as
\begin{equation}
R_{m}(r_{j}, r_{j-1}; \tau )=
\frac{M}{i\hbar \varepsilon
}\,\exp\left\{\frac{i}{\hbar}\left[\frac{M}{2\varepsilon }(r_{j}^{2} +
r_{j-1}^{2}) - V(r_{j})\varepsilon - V_{eff}(r_{j}) \varepsilon
\right]\right\}\,{\bf I}_{m/\sigma }^{1}(\hat{r}_{j}). \label{SRK2}
\end{equation}
This means that the radial path integral in a conical space can be
understood as a flat space path integral modified with the
effective potential (\ref{effV}). Furthermore, the integral angular
momentum $m$ of each partial wave is replaced by a non-integral angular
momentum $m/\sigma $. Note that $m/\sigma $ is not assumed to
be an integer or a fractional number but is yet countable. The angular
summation (\ref{KR}) for the full propagator is still over $m \in {\bf
Z}$.

The appearance of the effective potential $V_{eff}(r)$ in (\ref{SRK2})
as an addition to the assumed external potential $V(r)$ is remarkable.
What is more remarkable is however that (\ref{effV}) is identifiable with the
effect of the curvature of the conical surface imbedded in Euclidean
space. The mean curvature and the Gaussian curvature of the cone are given by
\begin{equation}
H(r) = \frac{\sqrt{1-\sigma ^{2}}}{2\sigma r} ~~~\mbox{and}
~~~K(r)=2\pi\frac{1-\sigma}{\sigma}\delta^{(2)}(x,y),
\end{equation}
respectively, where $\delta^{(2)}(x,y)$ represents the two-dimensional $\delta$-function (see \cite{IJR} for detail). Apparently the effective
potential (\ref{effV}) can be expressed as
\begin{equation}
V_{eff}(r) = - \frac{\hbar^{2}}{2M}H^{2}(r). \label{effV2}
\end{equation}
In the presence of a repulsive potential which regularizes the
singularity at $r=0$, i.e.\ propagator and wave function vanish at $r=0$, the Gaussian curvature has no effect and we can put (\ref{effV2}) into the form
\begin{equation}
V_{eff}(r) = \frac{\hbar^{2}}{2M}\left\{K(r) - H^{2}(r)\right\}.
\end{equation}
The last expression is indeed identical in form with the curvature term
(\ref{VC}) in the Schr\"odinger equation. This means that the path
integral calculation in conical space is consistent only with the
Schr\"odinger equation modified by the curvature term of the form of
(\ref{VC}).

\section{The energy spectrum and the wave functions}

In the above, we have expressed the short-time radial propagator in the
form (\ref{SRK2}) in order to emphasize the emergence of the effective
potential which is identifiable with the mean curvature of the conical
surface. Since the effective potential as is given by (\ref{effV}) is of
the inverse square form $\sim 1/r^{2}$, it can be absorbed via the
asymptotic recombination into the index of the modified Bessel function,
so that the function (\ref{bfI2}) can be given by
\begin{equation}
{\bf I}_{m}^{\sigma }(\hat{r}_{j}) \sim I_{\mu
(m)}\left(\frac{M}{i\hbar\varepsilon }\hat{r}_{j}^2\right)
\end{equation}
with
\begin{equation}
\mu(m)=\frac{1}{2\sigma }\sqrt{4m^{2} + \sigma ^{2} - 1}.
\end{equation}
Thus the short-time radial propagator may also be written as
\begin{equation}
R_{m}(r_{j}, r_{j-1}; \varepsilon )= \frac{M}{i\hbar\varepsilon }
\exp\left\{\frac{i}{\hbar}\left[\frac{M}{2\varepsilon }(r_{j}^{2} +
r_{j-1}^{2}) - V(r_{j})\varepsilon \right]\right\}
I_{\mu }\left(\frac{M}{i\hbar \varepsilon }\hat{r}_{j}^{2}\right).
\label{SRK3}
\end{equation}
Here the cone topology is completely encoded in the index of the modified
Bessel function. In other words, the short-time radial propagator
(\ref{SRK3}) is formally identical with that of flat space with angular
momentum $m \in {\bf Z}$ replaced by an effective angular momentum $\mu
(m) \in {\bf R}$. It is important to notice that whenever the centrally
symmetric problem in flat space is soluble by path integration the
corresponding problem on the conical space can also be solved by path
integration.

To carry out the radial path integration, we have to specify the
external potential $V(r)$. For convenience to our discussion, we consider
a combination of the harmonic oscillator potential and a repulsive
inverse-square potential,
\begin{equation}
V(r) = \frac{1}{2}M\omega ^{2}r^{2} + \frac{\kappa \hbar^{2}}{8\sigma
^{2}Mr^{2}}, \label{a-r}
\end{equation}
where $\kappa >1-\sigma ^{2} >0$. This represents long-range attraction
and short-range repulsion. The repulsive potential with the chosen
$\kappa $ removes the singularity at $r=0$. In fact the finite-time
radial path integral for the potential (\ref{a-r}) in flat space has
been explicitly evaluated \cite{PI,LI,IS,IJ}, the result being
\begin{equation}
R_{m}(r'', r'; \tau )=\frac{M\omega }{2\pi i\hbar \sin \omega \tau }
\exp\left\{\frac{iM\omega }{2\hbar}(r^{\prime 2} + r^{\prime\prime
2})\cot \omega \tau \right\}\,I_{\nu (m,1)}\left(\frac{M\omega }{i\hbar
\sin \omega \tau }r'r''\right)   \label{fRK}
\end{equation}
with
\begin{equation}
\nu (m,1) = \frac{1}{2}\sqrt{4m^{2} + \kappa}.
\end{equation}
As is mentioned above, the finite-time radial
propagator in the conical space can be immediately obtained from
(\ref{fRK}) by simply replacing $\nu (m,1)$ by
\begin{equation}
\nu (m, \sigma )=\frac{1}{2\sigma }\sqrt{4\nu ^{2}(m,1)+\sigma ^{2} -1}
=\frac{1}{2\sigma }\sqrt{4m^{2}+ \kappa + \sigma ^{2} -1}. \label{nu2}
\end{equation}
The desired result is therefore identical with (\ref{fRK}) if $\nu
(m, \sigma)$ of (\ref{nu2}) is in the place of $\nu (m, 1)$.

With the help of (\ref{KR}) and the Hille-Hardy formula \cite{GR},
the full propagator can be expressed in the spectral representation,
\begin{equation}
K({\bf r}^{\prime \prime}, {\bf r}^{\prime}; \tau )=
\sum_{n=0}^{\infty }\sum_{m=-\infty }^{\infty } e^{-(i\tau /\hbar)E_{nm}}
\Psi _{nm}^{\ast}(r^{\prime}, \theta^{\prime})
\Psi _{nm}(r^{\prime \prime}, \theta^{\prime \prime})
\end{equation}
with the energy spectrum and the energy eigenfunctions given,
respectively, by
\begin{equation}
E_{nm}=\hbar\omega \left(2n + 1 + \frac{1}{2\sigma }\sqrt{4m^{2} +
\kappa  + \sigma ^{2} -1} \right)  \label{spec}
\end{equation}
and
\begin{equation}
\Psi _{nm}(r, \theta )=N_{nm}e^{im\theta }r^{\nu(m,\sigma )}e^{-
(M\omega /2\hbar)r^{2}}\,_{1}F_{1}(-n, \nu (m,\sigma )+1; (M\omega
/\hbar)r^2)  \label{wf}
\end{equation}
where
\begin{equation}
N_{nm}=\frac{1}{\Gamma (\nu (m,\sigma )+1)}\sqrt{
\frac{\Gamma (n + \nu (m,\sigma ) +1)}{\pi
n!}}\left(\frac{M\omega}{\hbar}\right)^{(\nu +1)/2}.
\end{equation}
As we have assumed $\kappa \geq 1 - \sigma ^{2}$ for making the
system singularity-free, the energy eigenvalues (\ref{spec}) are assured
to be real. Since $\nu (m, \sigma )$ in (\ref{nu2}) also remains real,
the wave functions (\ref{wf}) vanish at $r=0$ for $\nu (m,\sigma )\neq
0$.

\section{Conclusion}

In the present paper path integration has been carried out explicitly
for a particle in the field of a topological defect characterized by the
conical metric (\ref{dl}). By the asymptotic recombination technique
applied to the short-time path integral, we have shown that the
quantum mechanics in a conical space is essentially identical to that in
flat space with rescaled angular momentum, and that the path
integration calculation naturally leads to an effective potential which
is proportional to the squared mean curvature of the conical space
imbedded in Euclidean space. We have seen that our path integral
calculation is consistent with the Schr\"odinger equation modified by
the Gaussian and mean curvatures as in \cite{JK}.


\begin{thebibliography}{99}

\bibitem{KF}M. Kleman and J. Friedel, Rev. Mod. Phys. {\bf 80} 61
(2008)
\bibitem{Pod} B. Podolsky, Phys. Rev. {\bf 32} 812 (1928)
\bibitem{BdW} B.S. DeWitt, Rev. Mod. Phys. {\bf 29} 337 (1957)
\bibitem{JK}H. Jensen and H. Koppe, Ann. Phys. {\bf 63} 589
(1971)
\bibitem{daC}R.C.T. da Costa, Phys. Rev. A {\bf 23} 1982 (1981)
\bibitem{DMO}C. Destri, P. Maraner and E. Onofri, Nuovo Cimento
A {\bf 107}, 1826 (1994)
\bibitem{M}P. Maraner, J. Phys. A {\bf 28} 2939 (1995)
\bibitem{FH}R. Froese and I. Herbst, Comm. Math. Phys. {\bf 220}
489 (2001)
\bibitem{FC}G. Ferrari and G. Cuoghi, Phys. Rev. Lett. {\bf 100}
230403 (2008)
\bibitem{IJR}A. Inomata, G. Junker and J. Raynolds, arXiv:1110.2044
\bibitem{PS} R.A. Puntigam and H.H. Soleng, Class. Quantum Grav.
{\bf 14} 1129 (1997)
\bibitem{Vil}A. Vilenkin,  Phys. Rev. D {\bf 23} 852 (1981)
\bibitem{DJt}S. Deser, R. Jackiw and G. 't Hooft, Ann. Phys. {\bf 152} 220 (1984)
\bibitem{DJ}S. Deser and R. Jackiw, Comm. Math. Phys. {\bf 118}
495 (1988)
\bibitem{KS}B.S. Kay and U.M. Studer, Comm. Math. Phys. {\bf 139}
103 (1991)
\bibitem{FM}C. Filgueiras and F. Moraes, Ann. Phys. {\bf 323}
3150 (2008)
\bibitem{Feyn}R.P. Feynman and A.R. Hibbs, {\sl Quantum Mechanics and Path
Integrals} (McGraw-Hill, 1965)
\bibitem{EG} S.F. Edwards and Y.V. Gulyaev, Proc. Roy. Soc. London A
{\bf 279} 229 (1964)
\bibitem{PI} D. Peak and A. Inomata, J. Math. Phys. {\bf 10} 1422 (1969)
\bibitem{BJ}M. B\"ohm and G. Junker, J. Math. Phys. {\bf 28} 1978 (1987)
\bibitem{IKG} A. Inomata, H. Kuratsuji and C.C. Gerry, {\sl Path
Intgrals and Coherent States of SU(2) and SU(1,1)} (World Scientific,
Singapore, 1992)
\bibitem{LI}W. Langguth and A. Inomata, J. Math. Phys. {\bf 20} 499
(1979)
\bibitem{IS} A. Inomata and V.A. Singh, J. Math. Phys. {\bf 19} 2318
(1978)
\bibitem{IJ} A. Inomata and G. Junker, in {\sl Noncompact Lie Groups and
Some of Their Applications}, eds. E.A. Tanner and R. Wilson (Kluwer,
Dordrecht, 1994) p. 199.
\bibitem{GR} I.S. Gradshteyn and I.W. Ryzhik, {\sl Table of Integrals,
Series and Products} (Academic Press, New York, 1965) p.1038.
\end{thebibliography}
\end{document}